\documentclass[prl,twocolumn,superscriptaddress]{revtex4}
\usepackage{amssymb,amsmath,amsthm,graphicx}
\usepackage{latexsym}
\usepackage{bm}
\usepackage[]{hyperref}
\usepackage{cleveref}

\allowdisplaybreaks

\pdfoutput=1

\begin{document}

\title{Collapse and Nonlinear Instability of AdS with Angular Momenta}

\author{Matthew~W.~Choptuik}
\email{choptuik@physics.ubc.ca}
\affiliation{Department of Physics and Astronomy, University of British Columbia, 6224 Agricultural Road, Vancouver, B.C., V6T 1W9, Canada} 
\affiliation{CIFAR Cosmology and Gravity Program, 180 Dundas St W, Suite 1400, Toronto, Ontario, M5G 1Z8 Canada}
\author{\'Oscar~J.~C.~Dias}
\email{O.J.Campos-Dias@soton.ac.uk}
\affiliation{STAG research centre and Mathematical Sciences, University of Southampton, UK} 
\author{Jorge~E.~Santos}
\email{jss55@cam.ac.uk}
\affiliation{Department of Applied Mathematics and Theoretical Physics, University of Cambridge, Wilberforce Road, Cambridge CB3 0WA, UK} 
\author{Benson~Way}
\email{benson@phas.ubc.ca}
\affiliation{Department of Physics and Astronomy, University of British Columbia, 6224 Agricultural Road, Vancouver, B.C., V6T 1W9, Canada} 

\begin{abstract}
We present a numerical study of rotational dynamics in AdS$_5$ with equal angular momenta in the presence of a complex doublet scalar field. We determine that the endpoint of gravitational collapse is a Myers-Perry black hole for high energies and a hairy black hole for low energies. We investigate the timescale for collapse at low energies $E$, keeping the angular momenta $J\propto E$ in AdS length units. We find that the inclusion of angular momenta delays the collapse time, but retains a $t\sim1/E$ scaling.  We perturb and evolve rotating boson stars, and find that boson stars near AdS appear stable, but those sufficiently far from AdS are unstable.  We find that the dynamics of the boson star instability depend on the perturbation, resulting either in collapse to a Myers-Perry black hole, or development towards a stable oscillating solution. 
\end{abstract}

\maketitle

{\bf~Introduction --} Spacetimes with anti-de Sitter (AdS) boundary conditions play a central role in our understanding of gauge/gravity duality \cite{Maldacena:1997re,Gubser:1998bc,Witten:1998qj,Aharony:1999ti}, where solutions to the Einstein equation with a negative cosmological constant are dual to states of strongly coupled field theories. This correspondence has inspired the study of gravitational physics in AdS over the past two decades.

It is perhaps surprising that the issue of the nonlinear stability of (global) AdS was only raised nine years after AdS/CFT was first formulated \cite{DafermosHolzegel2006,Dafermos2006}.  Dafermos and Holzegel conjectured a nonlinear instability where the reflecting boundary of AdS allows for small but finite energy perturbations to grow and eventually collapse into a black hole. This is in stark contrast with Minkowksi and de Sitter spacetimes, where nonlinear stability has long been established \cite{friedrich86,Christodoulou:1993uv}.

The first numerical evidence in favour of such an instability of AdS was reported in \cite{Bizon:2011gg}. This topic has since attracted much attention both from numerical and formal perspectives \cite{Dias:2011ss,Dias:2012tq,Buchel:2012uh,Buchel:2013uba,Maliborski:2013jca,Bizon:2013xha,Maliborski:2012gx,Maliborski:2013ula,Baier:2013gsa,Jalmuzna:2013rwa,Basu:2012gg,2012arXiv1212.1907G,Fodor:2013lza,Friedrich:2014raa,Bizon2014,Maliborski:2014rma,Abajo-Arrastia:2014fma,Balasubramanian:2014cja,Bizon:2014bya,Balasubramanian:2015uua,daSilva:2014zva,Craps:2014vaa,Basu:2014sia,Deppe:2014oua,Dimitrakopoulos:2014ada,Horowitz:2014hja,Buchel:2014xwa,Craps:2014jwa,Basu:2015efa,Yang:2015jha,Fodor:2015eia,Okawa:2015xma,Bizon:2015pfa,Dimitrakopoulos:2015pwa,Green:2015dsa,Deppe:2015qsa,Craps:2015iia,Craps:2015xya,Evnin:2015gma,Menon:2015oda,Jalmuzna:2015hoa,Evnin:2015wyi,Freivogel:2015wib,Dias:2016ewl,Evnin:2016mjx,Deppe:2016gur,Dimitrakopoulos:2016tss,Dimitrakopoulos:2016euh,Rostworowski:2016isb,Jalmuzna:2017mpa,Rostworowski:2017tcx,Martinon:2017uyo,Moschidis:2017lcr,Moschidis:2017llu,Dias:2017tjg}. Remarkably, this instability has recently been proved for the spherically symmetric and pressureless Einstein-massless Vlasov system \cite{Moschidis:2017lcr,Moschidis:2017llu}.

The collapse timescale is dual to the thermalisation time in the field theory, and is important for characterising and understanding this instability.  For energies $E$ much smaller than the AdS length $L=1$, early evolution is well-described by perturbation theory.  However, irremovable resonances generically cause secular terms to grow, leading to a breakdown of perturbation theory at a time $t\sim 1/E$. Numerical evidence suggests that horizon formation occurs shortly thereafter, \textit{i.e.}~at this same timescale.  It is not fully understood why collapse seems to occur at the shortest timescale allowed by perturbation theory, though see \cite{Freivogel:2015wib} for some recent progress.

However, all numerical studies have been restricted to zero angular momentum. Though perturbation theory breaks down at $t\sim 1/E$ for systems with rotation as well \cite{Dias:2011ss,Dias:2016ewl,Dias:2017tjg}, this only places a lower bound on the timescale for gravitational collapse. It therefore remains unclear whether rotational forces could balance the gravitational attraction and delay the collapse time.  

The inclusion of angular momentum also enriches the phase diagram of solutions. In addition to the Myers-Perry (and Kerr) family of rotating black holes, there are ``black resonators" \cite{Dias:2015rxy} which can be described as black holes with gravitational hair, and ``geons" \cite{Dias:2011ss,Horowitz:2014hja,Martinon:2017uyo} which are horizonless gravitational configurations held together by their own self-gravity.  The nonlinear dynamics of these solutions remain largely unexplored.

Due to the lack of symmetries, the inclusion of angular momenta poses a numerical challenge (though see \cite{Bantilan:2017kok} for recent progress away from spherical symmetry). For instance, the dynamical problem for pure gravity in four dimensions requires a full 3+1 simulation. To reduce numerical cost (see \cite{Bizon:2005cp,Bizon2014,Jalmuzna:2015hoa,Jalmuzna:2017mpa} for the use of a similar strategy), we will rely on the fact that in odd dimensions $(d\geq5)$, black holes have an enhanced symmetry when all of their angular momenta are equal. This simplification alone is insufficient for our purposes since gravitons that carry angular momenta break these symmetries. We therefore introduce a complex scalar doublet $\Pi$ given by the action
\begin{equation}\label{action}
S=\frac{1}{16\pi G_5}\int\mathrm d^5x\sqrt{-g}\left(R+12-2|\nabla\Pi|^2\right)\;.
\end{equation}
As we shall see, this theory admits an ansatz with which one can study gravitational collapse with angular momentum using a 1+1 numerical simulation. 

Moreover, this ansatz has a phase diagram of stationary solutions that is similar to that of pure gravity \cite{Dias:2011at}. In particular, this theory contains hairy black holes and boson stars, which are somewhat analogous to black resonators and geons, respectively.  Consequently, in addition to gravitational collapse, we are also able to investigate the dynamics of hairy black holes and boson stars.

Again, in this context, hairy black holes are much like black resonators, only with scalar hair instead of gravitational hair.  Both the hairy black holes and boson stars exist for energies and angular momenta where Myers-Perry black holes are super-extremal and singular. For these conserved quantities, the weak cosmic censorship conjecture therefore implies that the final state following gravitational collapse cannot be a Myers-Perry black hole.  For evolution respecting the symmetries of our ansatz, we wish to test cosmic censorship by identifying the endpoint of collapse.

Boson stars are horizonless solutions with a stationary metric and harmonically oscillating scalar field \cite{Dias:2011at,Liebling:2012fv,Buchel:2013uba}.  They are important objects for the study of the AdS instability since, like geons (and oscillons \cite{Maliborski:2013jca,Fodor:2015eia} for a real scalar field), they can be generated as nonlinear extensions of normal modes of AdS.  Such solutions can avoid the resonance phenomenon that leads to perturbative breakdown. Indeed, simulations of some of these solutions indicate stability well past $t\sim 1/E$.  Initial data near these solutions therefore lie within an ``island of stability".  We wish to investigate whether this stability applies for rotating boson stars.

However, boson stars far from AdS (\textit{i.e.}, past a turning point in their phase diagram) are expected to be unstable.  We aim to determine the endpoint of this instability.

{\bf~Method --}  We describe our ansatz and equations of motion schematically here; 
a full account is given in the Appendix. We take our metric and scalar to be
\begin{subequations}
\begin{align}\label{metricansatz}
\mathrm ds^2&=\frac{1}{(1-\rho^2)^2}\bigg\{-\alpha^2\left[1-\rho^2(2-\rho^2)\frac{\beta^2}{a}\right]\mathrm dt^2+\nonumber\\
&\qquad+\frac{4\alpha\beta}{a}\rho\,\mathrm dt\mathrm d\rho+\frac{\mathrm d\rho^2}{a(2-\rho^2)}+\nonumber\\
&\qquad+\rho^2(2-\rho^2)\bigg[\frac{1}{b^2}\Big(\mathrm d\psi+\cos^2(\tfrac{\theta}{2})\mathrm d\phi-\Omega \mathrm dt\Big)^2+\nonumber\\
&\qquad\qquad\qquad\qquad+\frac{b}{4}\Big(\mathrm d\theta^2+\sin^2\theta\,\mathrm d\phi^2\Big)\bigg]\bigg\}\;,\\
\Pi&=(\Pi_\mathfrak R+i \,\Pi_\mathfrak I)\begin{bmatrix}
       &e^{i\psi}\sin(\frac{\theta}{2})& \\
       &e^{i(\psi+\phi)}\cos(\frac{\theta}{2})& \\
\end{bmatrix}\;,
\end{align}
\end{subequations}
where $\alpha$, $\beta$, $a$, $\Omega$, $b$, $\Pi_\mathfrak R$, and $\Pi_\mathfrak I$, are real functions of $t$ and $\rho$ only. This ansatz has $SU(2)$ rather than $U(1)\times U(1)$ symmetry due to the equal rotation in the $\psi$ and $\psi+\phi$ angles in orthogonal planes.  This symmetry is preserved by $\Pi$.  Without the scalar field, one finds that horizonless solutions have $\Omega=0$ and hence do not rotate. 

Gauge freedom is fixed by maximal slicing, where the trace of the extrinsic curvature $K=0$ \cite{Choptuik:1999gh}.  Unlike the choice $\beta=0$ in other studies, this gauge allows for evolution beyond the formation of an apparent horizon. 

To bring the equations to first-order form, we introduce $w$ and $\delta$ as functions related directly to $\partial_\rho\Omega$ and $\partial_\rho\alpha$, respectively.  Let the vector $\varphi=\{b,\Pi_\mathfrak R,\Pi_\mathfrak I\}$, and introduce the vectors $p$, $q$ related to $\partial_t\varphi$ and $\partial_\rho\varphi$, respectively.  Finally, define $u=\{w,\beta,a\}$ and $v=\{\delta,\alpha,\Omega\}$.

The vectors $\varphi$, $p$, $q$, $u$, $v$ represent our 15 dynamical functions. Their equations of motion can be summarized as follows. First, the definition of $q$ takes the form
\begin{equation}\label{constraint}
\partial_\rho\varphi+A_\varphi[\rho,a]\varphi=g_\varphi[\rho,a,q]
\end{equation}
where $A_\varphi$ is a matrix and $g_\varphi$ is a vector.

Second, we have the evolution equations
\begin{equation}\label{evol}
\begin{bmatrix}
       \partial_t\varphi \\
       \partial_tq \\
       \partial_tp \\
\end{bmatrix}
=
\begin{bmatrix}
       0&0&0 \\
       \gamma A_\gamma&A_d&A_q \\
       0&A_p&A_d \\
\end{bmatrix}
\begin{bmatrix}
       \partial_\rho\varphi \\
       \partial_\rho q \\
       \partial_\rho p \\
\end{bmatrix}
+
\begin{bmatrix}
       f_\varphi \\
       f_q+\gamma f_\gamma \\
       f_p \\
\end{bmatrix}
\;,
\end{equation}
where the $A$'s are matrices that do not depend on $\varphi$, $q$, or $p$, and the $f$'s are vector-valued nonlinear expressions that can depend on all the dynamical
functions. The first row of the equation above includes the definition of $p$. We have incorporated \eqref{constraint} into these evolution equations as damping terms with $\gamma$ acting as a damping coefficient.

Next, we have the slicing equations for $u$ and $v$
\begin{subequations}\label{slice}
\begin{align}
A_u[\rho] \partial_\rho u +B_u[\rho,\varphi,p]u&=g_u[\rho,\varphi,q,p,u]\;,\\
A_v[\rho] \partial_\rho v +B_v[\rho,\varphi,q,p,u]v&=g_v[\rho,\varphi,q,p,u,v]\;,
\end{align}
\end{subequations}
where the $A$'s and $B$'s are matrices and the $g$'s are vectors. Included within these equations are the definitions of $w$ and $\delta$. These are nonlinear systems due to the dependence of the $g$'s on $u$ and $v$. However, given $\varphi$, $q$, and $p$, one can find $u$ and $v$ by solving a sequence of linear problems, with the solution to one linear problem entering nonlinearly as the source term for the next. (See Appendix for details.)  Obtaining this property in the equations partially motivated the metric ansatz \eqref{metricansatz}.

Finally, we have the evolution equations for $u$
\begin{equation}\label{uevol}
\partial_t u=f_u\;,
\end{equation}
where $f_u$ is a vector that depends on all the functions.

Initial data is supplied as a choice of $\varphi$ and $p$. The remaining functions can be obtained by solving the nonlinear systems \eqref{constraint} and \eqref{slice} with Newton-Raphson iteration. We evolve the system with a fourth order Runge-Kutta method that steps $\varphi$, $p$, and $q$ through \eqref{evol}, and then obtains $u$ and $v$ by solving \eqref{slice} as a sequence of linear problems.  We compute expansion coefficients from the metric to determine if a horizon has formed. We also monitor \eqref{uevol} and \eqref{constraint} as a check on numerics.

At infinity ($\rho=1$), we fix the boundary metric to be that of global AdS and require $\Pi=0$.  The energy $E$ and angular momentum $J$ are read from the metric at infinity, and are conserved by \eqref{uevol}.  The response of the scalar field $\langle\Pi\rangle$ is obtained by
\begin{equation}
\Pi=(1-\rho^2)^4\langle\Pi\rangle+\mathcal O[(1-\rho^2)^5]\;.
\end{equation}

Prior to horizon formation, we require regularity at the origin $\rho=0$.  After horizon formation, our numerical grid will be excised, and boundary conditions at the excision surface are supplied for $u$ through \eqref{uevol}, and the value of $\delta$ is held fixed~\footnote{A choice of $\delta$ at the excision surface fixes residual gauge freedom.}.  No other boundary conditions are required at the excision surface.

We use a spectral element mesh with Legendre-Gauss-Lobatto nodes, and inter-element coupling handled by a discontinuous Galerkin method with Lax-Friedrichs flux. Adaptive mesh refinement for splitting elements and increasing/decreasing polynomial order is decided by monitoring the Legendre spectrum within each element.  Linear systems are solved via sparse LU decomposition.  For all data presented here, relative energy and momentum violation, and violation of \eqref{constraint} are within or below $10^{-8}$. See the Appendix for more on numerical checks. 

{\bf~Gaussian Data --} Consider Gaussian initial data
\begin{subequations}
\begin{align}\label{gaussiandata}
\Pi_\mathfrak I |_{t=0}&= \epsilon \,\rho \,\sqrt{2-\rho^2} \,(1-\rho^2)^4 \,e^{-4[2(1-\rho^2)-1]^2}\;,\\
\frac{\partial_t\Pi_\mathfrak R}{\alpha\sqrt{a}}\bigg|_{t=0}&=4\,\lambda \,\Pi_\mathfrak I
\end{align}
\end{subequations}
with the other functions within $\varphi$ and $p$ vanishing.  This data is parametrised by $\epsilon$ and $\lambda$. At fixed $\lambda$ and small varying $\epsilon$, we have $E\propto\epsilon^2$ and $J\propto E$.  At larger $\epsilon$, deviation from this scaling may occur. This is a natural choice of parameters since individual normal modes of AdS that carry angular momentum also obey $J\propto E$ at small $E$. We take two families of initial data: one with fixed $\lambda=0$ where $J=0$, and another with $\lambda=1$, where $J\approx 0.155 E$. 

Let us describe the stationary black holes that can serve as final states of gravitational collapse.  These black holes must fall within the symmetry class of our ansatz and have the same conserved quantities as our initial data.  For $\lambda=0$, we have $J=0$ so the only stationary black holes are Schwarzschild-Tangherlini solutions.  

For $\lambda=1$, there are two competing families of regular black hole solutions. Myers-Perry black holes have the most entropy where they exist, \textit{i.e.} for $E>E_\mathrm{extr}\approx0.0691$. For all energies $E<E_\mathrm{extr}$, hairy black holes have the most entropy (by being the only existing solution).  We wish to verify that gravitational collapse for the $\lambda=1$ family will eventually settle into one of these black holes in accord with their respective energy ranges. This can be viewed as a test of cosmic censorship, since Myers-Perry would be superextremal for $E<E_\mathrm{extr}$.  

\begin{figure}
\centering
\includegraphics[width=.47\textwidth]{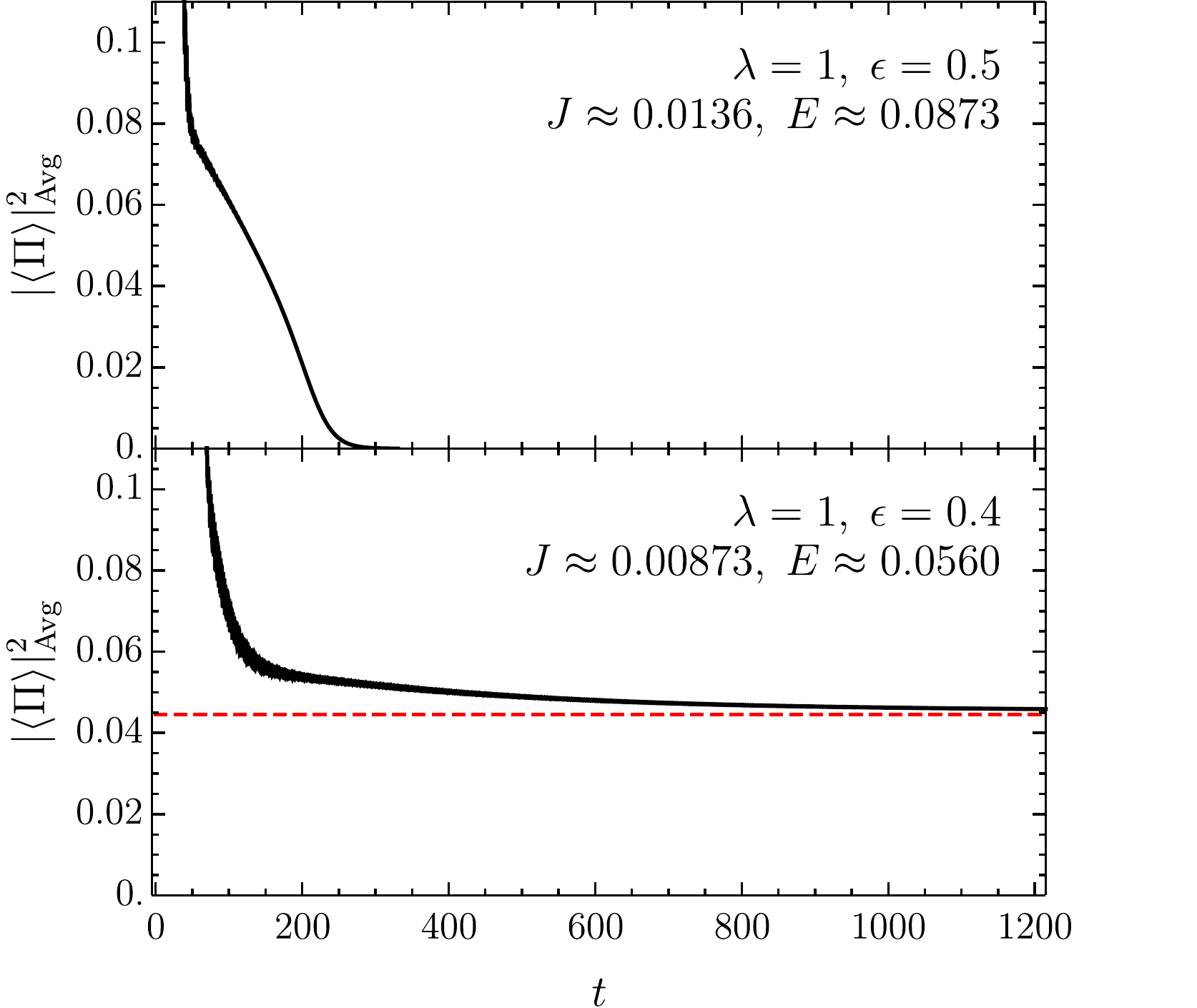}
\caption{Evolution of scalar response, averaged in a $2\pi$ time window, for Gaussian initial data. \textit{Top:} Collapse occurs at $t\approx30.1$ and settles to a Myers-Perry black hole. \textit{Bottom:} Collapse occurs at $t\approx55.4$ and settles to a hairy black hole. The red dashed line is the hairy black hole value from \cite{Dias:2011at}.}\label{fig:finalstate}
\end{figure}  

A hairy black hole can be distinguished from the Myers-Perry case by the presence of the scalar field.  In Fig.~\ref{fig:finalstate}, we show the evolution of the normed square of the scalar response $|\langle\Pi\rangle|^2$ for two cases.  In the first case, $E\approx0.0873>E_\mathrm{extr}$, so the Myers-Perry configuration is the preferred solution. Indeed, we find that the scalar field vanishes at late times.  

In the second case, $E\approx0.0560< E_\mathrm{extr}$, so the Myers-Perry black hole is superextremal, and we find that the scalar field approaches a constant non-zero value at late times, indicative of a hairy black hole. We have also matched this value to that of the expected final hairy black hole solution which was first obtained in \cite{Dias:2011at}.

In both cases, we have also matched the final entropy and angular frequency to their respective final stationary solutions \cite{Dias:2011at}.  In the subextremal case, we have matched quasinormal modes as well. (See Appendix.)  

\begin{figure}
\centering
\includegraphics[width=.47\textwidth]{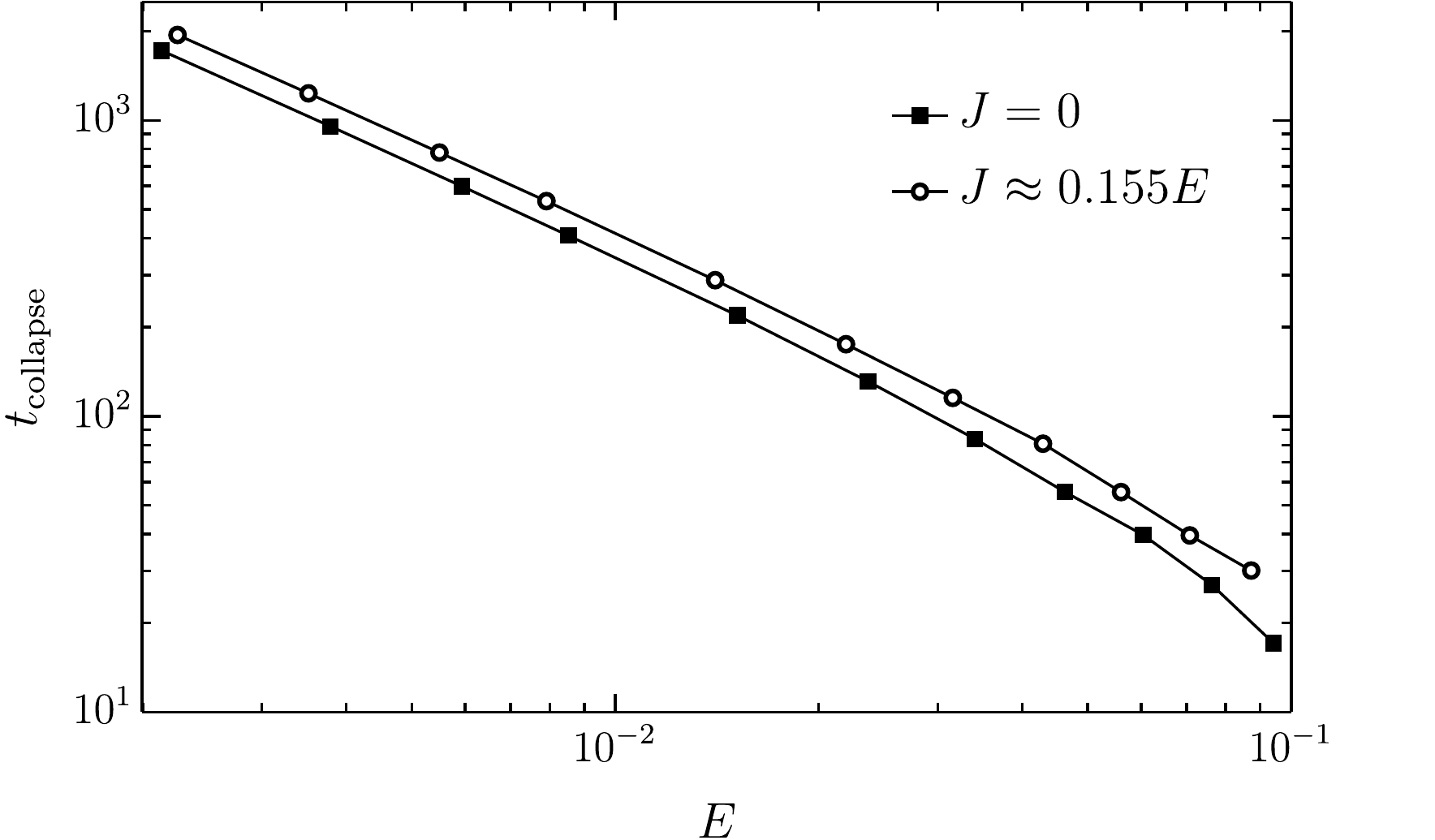}
\caption{Collapse times versus energy for Gaussian initial data. The power-law is consistent with a $t\sim1/E$ scaling. The two longest runs collapse at $t\approx 1721.12$ and $t\approx 1944.19$.}\label{fig:tcollapse}
\end{figure}  

Now we compare the timescale for horizon formation between the $\lambda=0$ ($J=0$) and $\lambda=1$ ($J\approx 0.155 E$) families of initial data.  In Fig.~\ref{fig:tcollapse}, we show a log-log plot of the collapse time versus the energy.  We see that at fixed energy, the initial data with nonzero angular momentum takes a longer time to collapse.  However, the collapse times for both sets of initial data exhibit a power-law that is consistent with a $t\sim1/E$ scaling.  We conclude that in this case, angular momentum increases the collapse time but does not affect the timescale. 

{\bf~Boson Star Data --} Boson stars within our ansatz have been constructed in \cite{Dias:2011at}, and can be found by setting the metric to be time-independent and the scalar field to have a harmonically oscillating complex phase.  They can be parametrised by their harmonic frequency $\omega$. For small energies, $\omega$ is close to a normal mode frequency of AdS. We focus on the lowest frequency mode with $\omega=5$ near AdS.  For small energies, these particular boson stars have angular momentum $J\simeq0.2E$.

As one increases the energy of the boson star, $\omega$ decreases until a turning point is reached around $\omega\approx 4.35$, where $E$ and $J$ are both maximal. Boson stars that lie on the AdS side of this turning point are expected to be nonlinearly stable (at least up to $t\sim 1/E$), and are otherwise expected to be unstable~\footnote{Morse theory typically implies that solutions on one side of a turning point are unstable. For completeness, we demonstrate linear instability in the Appendix.}. 

\begin{figure}
\centering
\includegraphics[width=.5\textwidth]{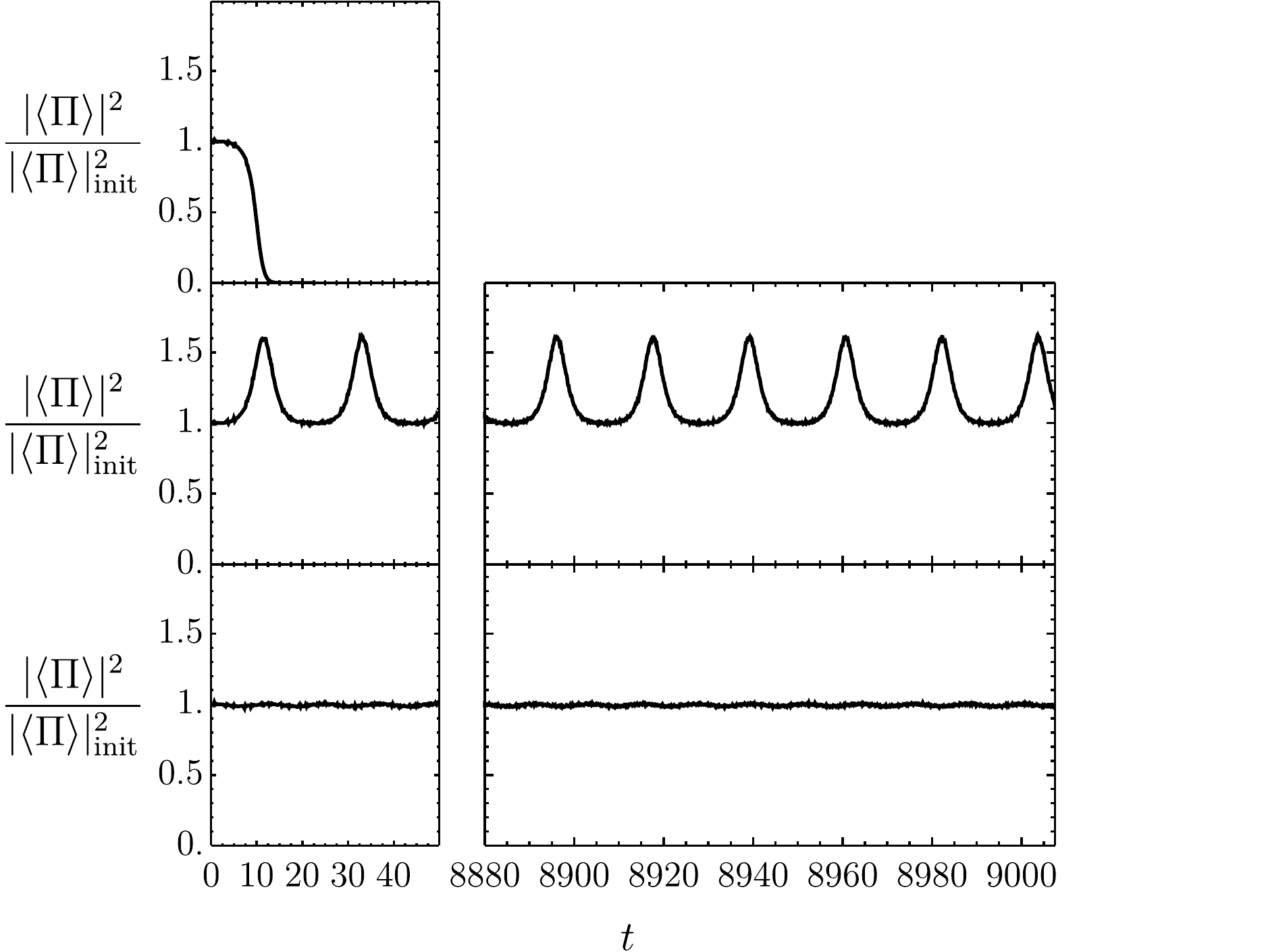}
\caption{Evolution of scalar response, scaled by its initial value, for perturbed boson stars near the turning point. \textit{Top:} Boson star with $\omega=4.3$ collapses into a Myers-Perry black hole.   \textit{Middle:} Boson star with $\omega=4.3$ evolves to a stable oscillon. The perturbations for the top and middle plots differ by a sign.  \textit{Bottom:} Boson star with $\omega=4.4$ remains stable.}\label{fig:bstars}
\end{figure}  

We perturb boson stars near the turning point with frequencies $\omega=4.3$ (in the `unstable' branch) and $\omega=4.4$ (in the `stable' branch) with a Gaussian profile similar to \eqref{gaussiandata}. Their scalar response $|\langle\Pi\rangle|^2$ is shown in Fig.~\ref{fig:bstars}.  Note that though the scalar field oscillates with frequency $\omega$, these oscillations are canceled out in $|\Pi|^2$, and consequently are not seen in Fig.~\ref{fig:bstars} nor in the metric.

Indeed, the $\omega=4.3$ boson star is unstable, but the endpoint of its evolution depends on the perturbation. For one perturbation (top panel of Fig.~\ref{fig:bstars}), evolution proceeds rapidly towards gravitational collapse, and eventually settles to a Myers-Perry black hole.  While a competing hairy black hole also exists, it has less entropy than Myers-Perry in this region of parameter space.

Perturbing the same boson star with the opposite sign yields drastically different results. As one can see from the middle panel of Fig.~\ref{fig:bstars}, large $O(1)$ deformations develop in $|\langle\Pi\rangle|^2$ (and the metric) that oscillate for long times. The frequency of these oscillations is much smaller than the boson star frequency $\omega$.  The metric and scalar both oscillate, so the final state (assuming continued stability) might be characterised as an oscillon. Since the frequency $\omega$ is still present in the scalar, this solution is, in a sense, a multi-frequency oscillon. 

In contrast to the above, the lower panel of Fig.~\ref{fig:bstars} shows that the perturbed boson with $\omega=4.4$ remains stable at long times, with no large deviations from the initial data.

We have repeated this study for different perturbations and boson stars, and also for oscillons (where $\Pi_\mathcal I=0$, see also \cite{Fodor:2015eia}). We find no qualitative difference to the above. 

{\bf~Conclusions --} 
Our numerical results suggest that much of our understanding of the instability of AdS carries over to situations with angular momenta as well.  In particular, for generic data, the timescale for gravitational collapse $t\sim1/E$ is preserved in the presence of rotation. Additionally, like oscillons in spherical symmetry, there are solutions that are nonlinear extensions of normal modes of AdS that are stable past $t\sim1/E$. 

We have also found that, depending on the perturbation, unstable boson stars will either collapse or oscillate. A comparison can be made to situations in flat space where the endpoint of unstable solutions can also depend upon the perturbation (see, \textit{e.g.} \cite{Lee:1988av,Seidel:1990jh,Lai:2007tj,Liebling:2012fv,Sanchis-Gual:2017bhw,Carracedo:2016qrf}). In flat space, energy and angular momentum can be carried away, and the non-collapsing evolution is well-approximated by a perturbation of a stable boson star, presumably settling towards a stable boson star at asymptotically late time.  By analogy, we suspect that the oscillating solution we find in AdS can be described as a nonlinear extension of a perturbed stable boson star. In AdS, however, there is a reflecting boundary which may cause the oscillations to persist indefinitely. 

Finally, let us comment on interesting regions of parameter space that we have not studied. There is a range of energies and angular momenta where hairy black holes have more entropy than Myers-Perry black holes.  This happens to be where Myers-Perry black holes are unstable to superradiance. In fact, hairy black holes branch off from Myers-Perry configurations precisely at the onset of this instability, for particular perturbations \cite{Dias:2011at}. 

This region is therefore a natural place to study the rotational superradiant instability \cite{Hawking:1999dp,Cardoso:2004hs,Kunduri:2006qa,Dias:2013sdc,Cardoso:2013pza,Dias:2015rxy,Niehoff:2015oga,Green:2015kur} for which little is known fully dynamically.  However, typical growth rates for this instability are around $10^{-5}$ \cite{Dias:2011at}, which requires a longer simulation than we can feasibly perform with our methods.  Furthermore, our ansatz implies that such a study will necessarily be incomplete.  High angular wavenumbers are expected to play an important role in this instability \cite{Cardoso:2013pza,Dias:2015rxy,Niehoff:2015oga,Green:2015kur}, but our ansatz is restricted to only the $m=1$ azimuthal wavenumbers. 

\newpage

{\bf~Acknowledgements --} We would like to thank Mihalis Dafermos, Gary Horowitz, Luis Lehner and Harvey Reall for reading an earlier version of the manuscript. The authors thankfully acknowledge the computer resources, technical expertise and assistance provided by CENTRA/IST. Some computations were performed at the cluster ``Baltasar-Sete-Sois" and supported by the H2020 ERC Consolidator Grant ``Matter and strong field gravity: New frontiers in Einstein's theory" grant agreement no. MaGRaTh-646597. Some computations were performed on the COSMOS Shared Memory system at DAMTP, University of Cambridge operated on behalf of the STFC DiRAC HPC Facility and funded by BIS National E-infrastructure capital grant ST/J005673/1 and STFC grants ST/H008586/1, ST/K00333X/1. O.J.C.D. is supported by the STFC Ernest Rutherford grants ST/K005391/1 and ST/M004147/1. BW is supported by NSERC.
\newpage
\onecolumngrid
\appendix
\setcounter{equation}{0}
\renewcommand{\theequation}{A.\arabic{equation}}
\renewcommand{\thefigure}{A.\arabic{figure}}
\setcounter{figure}{0}

\begin{center}  
{\large\bf Appendix} 
\end{center}

\begin{center}  
{\bf Equations of Motion} 
\end{center} 
Here, we describe the equations of motion in full.  The ansatz, as we have presented it, is reproduced here:
\begin{subequations}
\begin{align}\label{appendix:metricansatz}
\mathrm ds^2&=\frac{1}{(1-\rho^2)^2}\bigg\{-\alpha^2\left[1-\rho^2(2-\rho^2)\frac{\beta^2}{a}\right]\mathrm dt^2+\frac{4\alpha\beta}{a}\rho\,\mathrm dt\mathrm d\rho+\frac{\mathrm d\rho^2}{a(2-\rho^2)}+\nonumber\\
&\qquad\qquad\qquad\quad+\rho^2(2-\rho^2)\bigg[\frac{1}{b^2}\Big(\mathrm d\psi+\cos^2(\tfrac{\theta}{2})\mathrm d\phi-\Omega \mathrm dt\Big)^2+\frac{b}{4}\Big(\mathrm d\theta^2+\sin^2\theta\,\mathrm d\phi^2\Big)\bigg]\bigg\}\;,\\
\Pi&=(\Pi_\mathfrak R+i\,\Pi_\mathfrak I)\begin{bmatrix}
       &e^{i\psi}\sin(\frac{\theta}{2})& \\
       &e^{i(\psi+\phi)}\cos(\frac{\theta}{2})& \\
\end{bmatrix}\;.
\end{align}
\end{subequations}
We first perform a number of function redefinitions as follows:
\begin{subequations}
\begin{align}
b&=1-\rho^2(2-\rho^2)(1-\rho^2)^4\varphi_b\;,\label{appendix:varphibdef}\\
\Pi_\mathfrak R&=\rho\sqrt{2-\rho^2}(1-\rho^2)^4\varphi_r\;,\\
\Pi_\mathfrak I&=\rho\sqrt{2-\rho^2}(1-\rho^2)^4\varphi_i\;,\\
\beta&=\rho^2(2-\rho^2)(1-\rho^2)^4u_\beta\;,\\
a&=1-\rho^2(2-\rho^2)(1-\rho^2)^4u_a\;,\\
\alpha&=1-(1-\rho^2)^4v_\alpha\;,\\
\Omega&=(1-\rho^2)^4v_\Omega\;.
\end{align}
\end{subequations}
With these redefinitions, the finiteness of the new functions will ensure that the boundary conditions are satisfied. Vacuum global AdS is recovered  when all of the new functions vanish. 

From here, we introduce a number of functions that will put the equations of motion into first-order form.  These are $q_b$, $q_r$, $q_i$, $p_b$, $p_r$, $p_i$, $u_w$ and $v_\delta$. We will give the definitions of these functions when presenting the equations of motion. 

For ease of presentation, let us also define a number of auxiliary expressions:
\begin{subequations}
\begin{align}
P^2&\equiv \frac{3\rho^2(2-\rho^2)}{4b^2}p_b^2+p_r^2+p_i^2\;,\\
W^2&\equiv b^2u_w^2+24u_\beta^2\;,\\
[PW]^2&\equiv a P^2+\frac{1}{4}\rho^2(2-\rho^2)W^2\;\\
Q^2&\equiv\frac{3\rho^2(2-\rho^2)}{4b^2}\Big[\rho(2-\rho^2)q_b+2(1-\rho^2)^2\varphi_b\Big]^2+\nonumber\\
&\qquad\qquad+\Big[\rho(2-\rho^2)q_r+(1-\rho^2)^2\varphi_r\Big]^2+\Big[\rho(2-\rho^2)q_i+(1-\rho^2)^2\varphi_i\Big]^2\;,\\
[QP]&\equiv\frac{3\rho^2(2-\rho^2)}{4b^2}\Big[\rho(2-\rho^2)q_b+2(1-\rho^2)^2\varphi_b\Big]p_b+\nonumber\\
&\qquad\qquad+\Big[\rho(2-\rho^2)q_r+(1-\rho^2)^2\varphi_r\Big]p_r+\Big[\rho(2-\rho^2)q_i+(1-\rho^2)^2\varphi_i\Big]p_i\;,\\
\Phi^2&\equiv\frac{\rho^2(2-\rho^2)(1+2b+3b^2)}{b^4}\varphi_b^2+\frac{2+b^3}{b}(\varphi_r^2+\varphi_i^2)\;,\\
[Q\Phi]^2&\equiv Q^2+(1-\rho^2)^2\Phi^2\;,\\
A^2&\equiv a\left(\Big[3+\rho^2(2-\rho^2)\Big]u_a+\frac{2(1-\rho^2)v_\delta}{\sqrt a}\right)\;,\\
[B_{\delta\alpha}]&\equiv\frac{2\rho^2(2-\rho^2)(1-\rho^2)^2}{3\sqrt{a}}\bigg\{(1-\rho^2)^4\bigg(\Big[6+\rho^2(2-\rho^2)\Big][PW]^2+[Q\Phi]^2\bigg)+3\Big[3-\rho^2(2-\rho^2)\Big]u_a\bigg\}\;,\\
S_a&\equiv \frac{4}{3}(1-\rho^2)^3\bigg(\rho^2(2-\rho^2)\Big[P^2+\frac{1}{4}\rho^2(2-\rho^2)W^2\Big]+[Q\Phi]^2\bigg)\;,\\
S_\delta&=(1-\rho^2)^2\bigg\{\rho^2(2-\rho^2)\bigg(2+\rho^2(2-\rho^2)(1-\rho^2)^4\Big[3+\rho^2(2-\rho^2)\Big]u_a\bigg)[PW]^2-\nonumber\\
&\qquad\qquad\qquad\qquad-\bigg(2-\rho^2(2-\rho^2)(1-\rho^2)^4\Big[3+\rho^2(2-\rho^2)\Big]u_a\bigg)[Q\Phi]^2\bigg\}+\nonumber\\
&\qquad+\rho^2(2-\rho^2)\bigg\{1-(1-\rho^2)^2\bigg(3-\rho^2(2-\rho^2)\Big[1-2\rho^2(2-\rho^2)\Big]\bigg)u_a\bigg\}u_a\;.
\end{align}
\end{subequations}

Now we present the equations of motion. We have a total of 15 functions which can expressed as $\varphi=\{\varphi_b,\varphi_r,\varphi_i\}$, $q=\{q_b,q_r,q_i\}$, $p=\{p_b,p_r,p_i\}$, $u=\{u_w,u_\beta,u_a\}$, and $v=\{v_\delta,v_\alpha,v_\Omega\}$.  The full set of equations of motion are all generated from a combination of the Einstein equation, the Klein-Gordon equation for the scalar doublet, the maximal slicing gauge condition $K=0$, and definitions of first-order functions.  There are a total of 21 equations.  

The first three equations are linear slicing equations for $\varphi$ that define the functions in $q$:
\begin{subequations}\label{appendix:qdef}
\begin{align}
(1-\rho^2)\partial_\rho\varphi_b-8\rho\left[1+\frac{(1-\rho^2)^6u_a}{2\sqrt{a}(1+\sqrt{a})}\right]\varphi_b&=\frac{2q_b}{\sqrt{a}}\;,\\
(1-\rho^2)\partial_\rho\varphi_r-8\rho\left[1+\frac{(1-\rho^2)^6u_r}{4\sqrt{a}(1+\sqrt{a})}\right]\varphi_r&=\frac{2q_r}{\sqrt{a}}\;,\\
(1-\rho^2)\partial_\rho\varphi_i-8\rho\left[1+\frac{(1-\rho^2)^6u_i}{4\sqrt{a}(1+\sqrt{a})}\right]\varphi_i&=\frac{2q_i}{\sqrt{a}}\;.
\end{align}
\end{subequations}
Note that the coefficient of the derivative term vanishes only at $\rho=1$. There is therefore a natural direction of integration for these equations, which is from the boundary $\rho=1$ towards the origin $\rho=0$.  No external boundary conditions are required for these equations. 

The next nine equations are evolution equations for $\varphi$, $q$, and $p$. The evolution equations for $\varphi$ are given by the definition of $p$:
\begin{subequations}\label{appendix:varphievol}
\begin{align}
\partial_t\varphi_b&=\alpha\left(\sqrt{a}\frac{p_b}{1-\rho^2}+\frac{1}{\sqrt a}{\rho^2(2-\rho^2)(1-\rho^2)^3u_\beta\Big[\rho(2-\rho^2)q_b+2(1-\rho^2)^2\varphi_b\Big]}\right)\;,\\
\partial_t\varphi_r&=\alpha\left(\sqrt{a}\frac{p_r}{1-\rho^2}+\frac{1}{\sqrt a}{\rho^2(2-\rho^2)(1-\rho^2)^3u_\beta\Big[\rho(2-\rho^2)q_r+(1-\rho^2)^2\varphi_r\Big]}\right)+(1-\rho^2)^4v_\Omega\varphi_i\;,\\
\partial_t\varphi_i&=\alpha\left(\sqrt{a}\frac{p_i}{1-\rho^2}+\frac{1}{\sqrt a}{\rho^2(2-\rho^2)(1-\rho^2)^3u_\beta\Big[\rho(2-\rho^2)q_i+(1-\rho^2)^2\varphi_i\Big]}\right)-(1-\rho^2)^4v_\Omega\varphi_r\;.
\end{align}
\end{subequations}
Note that finiteness of $\varphi$ requires that $p=0$ at $\rho=1$.  We must enforce this condition in our numerical evolution. 

The evolution equations for $q$ come from the commutation of time and spatial derivatives:
\begin{subequations}\label{appendix:qevol}
\begin{align}
\partial_tq_b&=\alpha\bigg\{\frac{a}{2}\partial_\rho p_b+\frac{1}{2}\rho^3(2-\rho^2)^2(1-\rho^2)^4u_\beta\partial_\rho q_b-3\rho\frac{p_b}{1-\rho^2}-\nonumber\\
&\qquad\qquad-\rho(1-\rho^2)^3\left[\frac{1}{3}(1-\rho^2)^4\Big[\rho^2(2-\rho^2)[PW]^2+[Q\Phi]^2\Big]+\bigg(1-6\rho^2(2-\rho^2)-\frac{2(1-\rho^2)^2}{1+\sqrt a}\bigg)u_a\right]p_b-\nonumber\\
&\qquad\qquad-\frac{1}{2}\rho(1-\rho^2)^3u_\beta\bigg(\rho(2-\rho^2)^2(1+13\rho^2)q_b+4(1-\rho^2)^2\Big[4+\rho^2(2-\rho^2)\Big]\varphi_b\bigg)\bigg\}+\frac{1}{2}\rho(1-\rho^2)^3p_bA^2+\nonumber\\
&\qquad+\gamma\left[\frac{\sqrt a}{2}(1-\rho^2)\partial_\rho \varphi_b-4\rho\left(\sqrt a+\frac{(1-\rho^2)^6u_a}{2(1+\sqrt a)}\right)\varphi_b-q_b\right]\;,\\
\partial_tq_r&=\alpha\bigg\{\frac{a}{2}\partial_\rho p_r+\frac{1}{2}\rho^3(2-\rho^2)^2(1-\rho^2)^4u_\beta\partial_\rho q_r-3\rho\frac{p_r}{1-\rho^2}-\nonumber\\
&\qquad\qquad-\rho(1-\rho^2)^3\left[\frac{1}{3}(1-\rho^2)^4\Big[\rho^2(2-\rho^2)[PW]^2+[Q\Phi]^2\Big]-\bigg(5\rho^2(2-\rho^2)+\frac{(1-\rho^2)^2}{1+\sqrt a}\bigg)u_a\right]p_r-\nonumber\\
&\qquad\qquad-\rho(1-\rho^2)^3\bigg[u_\beta\bigg(\frac{1}{2}\rho(2-\rho^2)^2\Big[4+\rho^2(21-11\rho^2)\Big]q_r+(1-\rho^2)^2\Big[4+\rho^2(2-\rho^2)\Big]\varphi_r\bigg)+(1-\rho^2)b^2u_w\varphi_i\bigg]\bigg\}+\nonumber\\
&\qquad+(1-\rho^2)^3\left[\frac{\rho}{2} A^2p_r+(1-\rho^2)v_\Omega q_i\right]+\gamma\left[\frac{\sqrt a}{2}(1-\rho^2)\partial_\rho \varphi_r-4\rho\left(\sqrt a+\frac{(1-\rho^2)^6u_a}{4(1+\sqrt a)}\right)\varphi_r-q_r\right]\;,\\
\partial_tq_i&=\alpha\bigg\{\frac{a}{2}\partial_\rho p_i+\frac{1}{2}\rho^3(2-\rho^2)^2(1-\rho^2)^4u_\beta\partial_\rho q_i-3\rho\frac{p_i}{1-\rho^2}-\nonumber\\
&\qquad\qquad-\rho(1-\rho^2)^3\left[\frac{1}{3}(1-\rho^2)^4\Big[\rho^2(2-\rho^2)[PW]^2+[Q\Phi]^2\Big]-\bigg(5\rho^2(2-\rho^2)+\frac{(1-\rho^2)^2}{1+\sqrt a}\bigg)u_a\right]p_i-\nonumber\\
&\qquad\qquad-\rho(1-\rho^2)^3\bigg[u_\beta\bigg(\frac{1}{2}\rho(2-\rho^2)^2\Big[4+\rho^2(21-11\rho^2)\Big]q_i+(1-\rho^2)^2\Big[4+\rho^2(2-\rho^2)\Big]\varphi_i\bigg)-(1-\rho^2)b^2u_w\varphi_r\bigg]\bigg\}+\nonumber\\
&\qquad+(1-\rho^2)^3\left[\frac{\rho}{2} A^2p_i-(1-\rho^2)v_\Omega q_r\right]+\gamma\left[\frac{\sqrt a}{2}(1-\rho^2)\partial_\rho \varphi_i-4\rho\left(\sqrt a+\frac{(1-\rho^2)^6u_a}{4(1+\sqrt a)}\right)\varphi_i-q_i\right]\;,
\end{align}
\end{subequations}
where we have included damping terms with a non-negative constant $\gamma$ acting as a damping coefficient. Note that the $\gamma$ terms vanish when \eqref{appendix:qdef} is satisfied.  One can show that deviations away from \eqref{appendix:qdef} will decay exponentially in time with a decay rate proportional to $\gamma$. As in the evolution equations for $\varphi$ \eqref{appendix:varphievol}, these equations require $p=0$ at $\rho=1$. 

The evolution equations for $p$ are
\begin{subequations}\label{appendix:pevol}
\begin{align}
\partial_tp_b&=\alpha\bigg\{\frac{1}{2}(2-\rho^2)\partial_\rho q_b+\frac{1}{2}\rho^3(2-\rho^2)^2(1-\rho^2)^4u_\beta\partial_\rho p_b+\frac{(14-15\rho^2)}{2}\frac{q_b}{\rho}-4(1-\rho^2)\varphi_b+\nonumber\\
&\qquad\qquad+\rho^2(2-\rho^2)(1-\rho^2)^3\bigg(\frac{2}{3}(1-\rho^2)^4[QP]+2\Big[1-3\rho^2(2-\rho^2)\Big]u_\beta-\frac{\sqrt a}{b}p_b\bigg)p_b+\nonumber\\
&\qquad\qquad+\frac{(1-\rho^2)^3}{\sqrt a}\bigg[\frac{1}{3}\rho^2(2-\rho^2)b^3u_w^2+\frac{1}{b}\Big[\rho(2-\rho^2)q_b+2(1-\rho^2)^2\varphi_b\Big]^2+\nonumber\\
&\qquad\qquad\qquad\qquad\qquad+\frac{2(1-\rho^2)^2u_a}{1+\sqrt a}\bigg(\rho(2-\rho^2)q_b-2\Big[1+\rho^2(2-\rho^2)\varphi_b\Big]\bigg)-\nonumber\\
&\qquad\qquad\qquad\qquad\qquad-\frac{4}{3}(1-\rho^2)^2\left(\frac{2}{b^3}(1+2b+3b^2)\varphi_b+(1-\rho^2)^4(1+b+b^2)(\varphi_r^2+\varphi_i^2)\right)\varphi_b\bigg]\bigg\}+\nonumber\\
&\qquad+\frac{(1-\rho^2)^3}{2a}\Big[\rho^4(2-\rho^2)^2(1-\rho^2)^4u_\beta p_b+\rho(2-\rho^2)q_b+2(1-\rho^2)^2\varphi_b\Big]A^2\;,\\
\partial_tp_r&=\alpha\bigg\{\frac{1}{2}(2-\rho^2)\partial_\rho q_r+\frac{1}{2}\rho^3(2-\rho^2)^2(1-\rho^2)^4u_\beta\partial_\rho p_r+\frac{(10-11\rho^2)}{2}\frac{q_r}{\rho}-(1-\rho^2)\varphi_r+\nonumber\\
&\qquad\qquad+\rho^2(2-\rho^2)(1-\rho^2)^3\bigg(\frac{2}{3}(1-\rho^2)^4[QP]+\Big[1-5\rho^2(2-\rho^2)\Big]u_\beta\bigg)p_r+\nonumber\\
&\qquad\qquad+\frac{(1-\rho^2)^5}{\sqrt a}\bigg[\frac{u_a}{1+\sqrt a}\bigg(\rho(2-\rho^2)q_r-\Big[2-\rho^2(2-\rho^2)\Big]\varphi_r\bigg)-\rho^2(2-\rho^2)(1-\rho^2)^4\frac{2+b}{b}\varphi_b^2\varphi_r\bigg]\bigg\}+\nonumber\\
&\qquad+(1-\rho^2)^3\bigg(\frac{1}{2a}\Big[\rho^4(2-\rho^2)^2(1-\rho^2)^4u_\beta p_r+\rho(2-\rho^2)q_r+(1-\rho^2)^2\varphi_r\Big]A^2+(1-\rho^2)v_\Omega p_i\bigg)\;,\\
\nonumber\\
\nonumber\\
\nonumber\\
\nonumber\\
\nonumber\\
\nonumber\\
\partial_tp_i&=\alpha\bigg\{\frac{1}{2}(2-\rho^2)\partial_\rho q_i+\frac{1}{2}\rho^3(2-\rho^2)^2(1-\rho^2)^4u_\beta\partial_\rho p_i+\frac{(10-11\rho^2)}{2}\frac{q_i}{\rho}-(1-\rho^2)\varphi_i+\nonumber\\
&\qquad\qquad+\rho^2(2-\rho^2)(1-\rho^2)^3\bigg(\frac{2}{3}(1-\rho^2)^4[QP]+\Big[1-5\rho^2(2-\rho^2)\Big]u_\beta\bigg)p_i+\nonumber\\
&\qquad\qquad+\frac{(1-\rho^2)^5}{\sqrt a}\bigg[\frac{u_a}{1+\sqrt a}\bigg(\rho(2-\rho^2)q_i-\Big[2-\rho^2(2-\rho^2)\Big]\varphi_i\bigg)-\rho^2(2-\rho^2)(1-\rho^2)^4\frac{2+b}{b}\varphi_b^2\varphi_i\bigg]\bigg\}+\nonumber\\
&\qquad+(1-\rho^2)^3\bigg(\frac{1}{2a}\Big[\rho^4(2-\rho^2)^2(1-\rho^2)^4u_\beta p_i+\rho(2-\rho^2)q_i+(1-\rho^2)^2\varphi_i\Big]A^2-(1-\rho^2)v_\Omega p_r\bigg)\;,
\end{align}
\end{subequations}
Finiteness of $p$ requires that $q=0$ at $\rho=0$.  This condition is enforced during numerical evolution.

This collection of nine evolution equations \eqref{appendix:varphievol}, \eqref{appendix:qevol}, and \eqref{appendix:pevol} take the form of an advection system
\begin{equation}
\partial_t\mathbf u=\mathbf A\partial_\rho\mathbf u+\mathbf f\;,
\end{equation}
for some vector of functions $\mathbf u$, advection operator $\mathbf A$, and nonlinear terms $\mathbf f$. 
The advection operator $\mathbf A$ has eigenvalues
\begin{equation}
0,\quad \frac{\alpha}{2}\Big[\rho^3(2-\rho^2)^2(1-\rho^2)^4u_\beta\pm\sqrt{2-\rho^2}\sqrt{a}\Big]\;,
\end{equation}
each with a three-fold degeneracy.  These eigenvalues define the characteristics of this system. The two signs give velocities to the ingoing and outgoing characteristics.  A computation of expansion coefficients indicates that an apparent horizon forms when
\begin{equation}
1-\rho^6(2-\rho^2)^3(1-\rho^2)^8\frac{u_\beta^2}{a}<0\;,
\end{equation}
which also corresponds to a situation where the outgoing characteristics switch sign. These eigenvalues are also used in our implementation of spectral element methods to determine flux across elements.

The next six equations are slicing equations for $u$ and $v$:
\begin{subequations}\label{appendix:uvslice}
\begin{align}
\rho(2-\rho^2)\partial_\rho u_w+12(1-\rho^2)u_w&=-8(1-\rho^2)^4(\varphi_rp_i-\varphi_ip_r)\;,\label{appendix:wslice}\\
\rho(2-\rho^2)\partial_\rho u_\beta+12(1-\rho^2)u_\beta&=\frac{4}{3}(1-\rho^2)^3[QP]\;,\label{appendix:betaslice}\\
\rho(2-\rho^2)\partial_\rho u_a+\frac{4}{3}(1-\rho^2)\Big[6+\rho^4(2-\rho^2)^2(1-\rho^2)^6P^2\Big]u_a&=S_{a}\;,\label{appendix:aslice}\\
\rho(2-\rho^2)\partial_\rho u_\delta+8(1-\rho^2)u_\delta+[B_{\delta\alpha}]v_\alpha&=\frac{1}{\sqrt a}S_{\delta}\;,\label{appendix:deltaslice}\\
(1-\rho^2)\partial_\rho v_\alpha-8\rho v_\alpha+\frac{2\rho(1-\rho^2)}{\sqrt a}v_\delta&=-\rho\left[3+\rho^2(2-\rho^2)\right]u_a\label{appendix:deltadef}\;,\\
(1-\rho^2)\partial_\rho v_\Omega-8\rho v_\Omega&=-\frac{2\rho\alpha b^2}{\sqrt a}u_w\;,\label{appendix:wdef}
\end{align}
\end{subequations}
where the last two equations \eqref{appendix:deltadef} and \eqref{appendix:wdef} define $u_\delta$ and $u_w$, respectively.  Each of these equations has a derivative term with a coefficient that vanishes only at $\rho=0$ or $\rho=1$. These determine the direction of integration for these equations.  Namely, the first four are integrated from the interior out, and the last two are integrated from the boundary in. Prior to horizon formation, the origin is part of the numerical grid, and no external boundary conditions are required. After horizon formation, the numerical grid interior to the horizon will be excised, and boundary conditions will need to be supplied for the first four equations.

This set of slicing equations \eqref{appendix:uvslice} is a nonlinear system, but can be solved as a series of linear systems. Suppose $\varphi$, $q$, and $p$ are known at a particular time slice.  Then \eqref{appendix:wslice} and \eqref{appendix:betaslice} can each be solved independently as a linear differential equation, obtaining $u_w$ and $u_\beta$. These functions enter into the source term $S_a$, and $u_a$ can then be obtained by solving \eqref{appendix:aslice}.  Then $u_w$, $u_\beta$, and $u_a$ are placed in $[B_{\delta\alpha}]$ and $S_\delta$, and the coupled linear system \eqref{appendix:wslice} and \eqref{appendix:betaslice} can be solved for $v_\delta$ and $v_\alpha$. Finally, $u_w$ and $v_\alpha$ enter into \eqref{appendix:wdef} whose solution yields $v_\Omega$.

Note that instead of evolving $\varphi$ through \eqref{appendix:varphievol}, we could instead solve the slicing equation for $\varphi$ \eqref{appendix:qdef} along with the slicing equations \eqref{appendix:uvslice}.  But, it is not possible to solve both \eqref{appendix:qdef} and \eqref{appendix:uvslice} as a series of linear equations. We wish to avoid using nonlinear solvers during evolution, so we have opted to solve just \eqref{appendix:uvslice} and handle \eqref{appendix:qdef} using damping terms in evolution.  Note, however, that to obtain initial data, we do directly solve \eqref{appendix:qdef} and \eqref{appendix:uvslice} using Newton-Raphson iteration.

Finally, the remaining three equations are evolution equations for $u$:
\begin{subequations}\label{appendix:uevol}
\begin{align}
\rho \partial_t u_w&=-4(1-\rho^2)^4\alpha\left[\varphi_rq_i-\varphi_iq_r+\rho^3(2-\rho^2)(1-\rho^2)^4u_\beta(\varphi_rp_i-\varphi_ip_r)\right]\;,\label{appendix:wevol}\\
\rho^2(2-\rho^2)\partial_t u_\beta&=\frac{\alpha}{3}\bigg\{(1-\rho^2)^3\bigg(\rho^2(2-\rho^2)\Big[a P^2-\frac{1}{4}\rho^2(2-\rho^2)W^2\Big]+\Big[Q^2-(1-\rho^2)^2\Phi^2\Big]\bigg)+\nonumber\\
&\qquad\qquad\qquad\qquad\qquad\qquad\qquad\qquad\qquad\qquad\qquad\qquad\qquad+2\rho^4(2-\rho^2)^2(1-\rho^2)^7u_\beta[QP]\bigg\}-\nonumber\\
&\qquad-\frac{1}{2}(1-\rho^2)\bigg[1-(1-\rho^2)^2\bigg(\rho^2(2-\rho^2)\Big[3+\rho^2(2-\rho^2)\Big]u_a-2\Big[1+\rho^2(2-\rho^2)\Big]v_\alpha\bigg)\bigg]u_a-\sqrt{a}v_\delta\;,\label{appendix:betaevol}\\
\partial_t u_a&=(1-\rho^2)^3\bigg\{\frac{2\alpha}{3}\bigg[\rho^2(2-\rho^2)u_\beta\bigg((1-\rho^2)^4\Big[\rho^2(2-\rho^2)[PW]^2+[Q\Phi]^2\Big]-3\Big[1+\rho^2(2-\rho^2)\Big]u_a\bigg)+2a[QP]\bigg]\nonumber\\
&\qquad\qquad\quad\quad+\rho^2(2-\rho^2)u_\beta A^2\bigg\}\;.\label{appendix:aevol}
\end{align}
\end{subequations}
These are degenerate evolution equations. That is, they do not contain spatial derivatives (or equivalently, their advection operator has only zero eigenvalues). Since the energy $E\propto u_a(t,\rho=1)$ and angular momentum $J\propto u_w(t,\rho=1)$, the vanishing of the equations \eqref{appendix:wevol} and \eqref{appendix:aevol} at $\rho=1$ imply that $E$ and $J$ are conserved.

After excision, these equations supply boundary conditions for $u_w$, $u_\beta$, and $u_a$ in the first three slicing equations of \eqref{appendix:uvslice}.  The remaining boundary condition is determined by keeping $u_\delta$ fixed.  One can show that this last condition just fixes an integration constant in the maximal slicing gauge condition $K=0$. 

Lastly, for completeness, we express our Gaussian initial data in terms of the variables in this section:
\begin{subequations}
\begin{align}
\varphi_i&=\epsilon\,e^{-4[2(1-\rho^2)^2-1]^2}\;,\\
p_r&=4\lambda\epsilon\,(1-\rho^2)e^{-4[2(1-\rho^2)^2-1]^2}\;,
\end{align}
\end{subequations}
with the remaining functions in $\varphi$ and $p$ vanishing.  The remaining functions can be obtained by solving \eqref{appendix:qdef} and \eqref{appendix:uvslice} using Newton-Raphson iteration.

\begin{center}  
{\bf Numerical Validation} 
\end{center} 

In this section, we present a number of additional numerical checks.  Recall that we evolve the system by stepping $\varphi$, $q$, and $p$ in time using \eqref{appendix:varphievol}, \eqref{appendix:qevol}, and \eqref{appendix:pevol}, then obtain $u$ and $v$ by solving the slicing equations \eqref{appendix:uvslice}. There are two sets of equations that we do not solve directly: the slicing equations for $\varphi$ given in \eqref{appendix:qdef}, and the evolution equations for $u$ in \eqref{appendix:uevol}. We can use these equations for numerical checks. 

We verify \eqref{appendix:qdef} by solving it for $\varphi$ and comparing the result to what we obtained through our main evolution. We call the difference of these quantities $\Delta_\mathrm{Evol} \varphi$. Similarly, given the numerical functions at a time slice, we take a single time step of the left hand side of \eqref{appendix:uevol} using \eqref{appendix:uevol} directly, and compare that result to what we obtain using our main evolution code. We call the difference of these $\Delta_\mathrm{Evol} u$. Finally, we monitor the relative violation of energy and angular momentum which are quantities conserved though \eqref{appendix:uevol}.

\begin{figure}
\centering
\includegraphics[width=.45\textwidth]{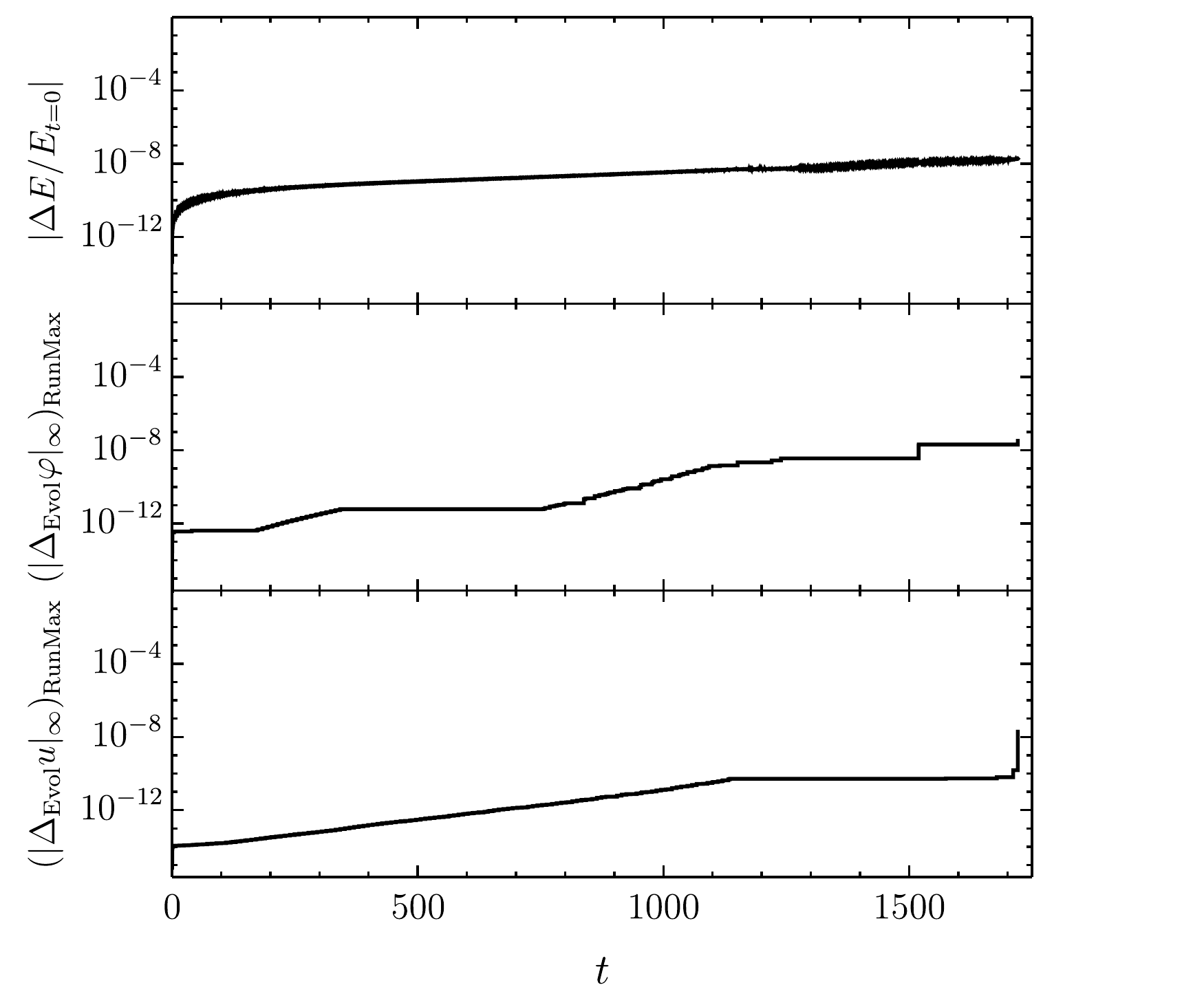}
\includegraphics[width=.45\textwidth]{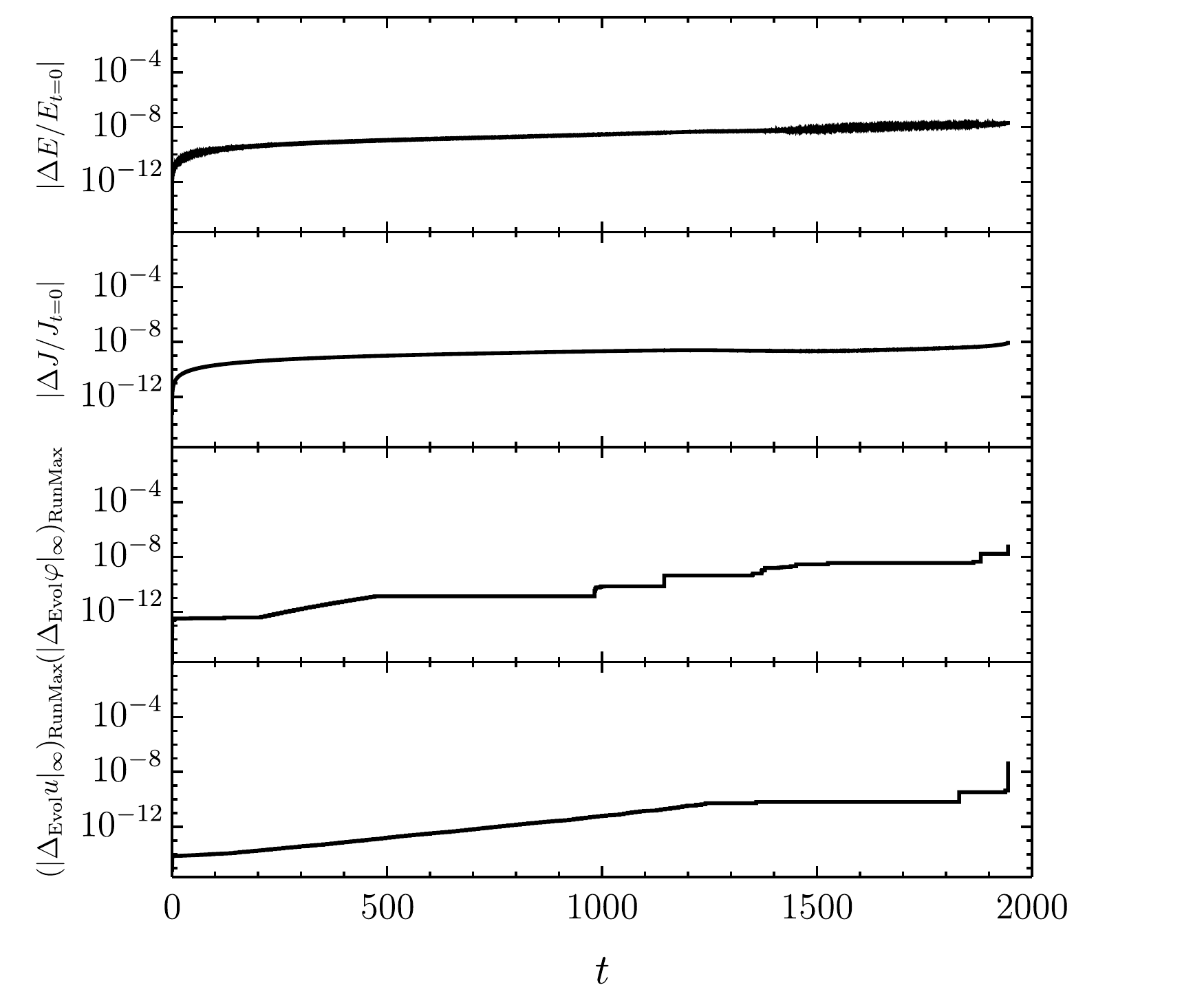}
\caption{Relative energy violation, momentum violation, and differences between our main evolution code 
and \eqref{appendix:qdef}, \eqref{appendix:uevol}.  For the latter two, we show the running maximum of the infinity norm. \textit{Left:} Our longest collapsing run with $\lambda=0$ (We do not show momentum violation since $J=0$). \textit{Right:} Our longest collapsing run with $\lambda=1$.}\label{appendixfig:error}
\end{figure}  

In Fig.~\ref{appendixfig:error}, we show these various measures of error as a function of time.  The differences $\Delta_\mathrm{Evol} \varphi$ and $\Delta_\mathrm{Evol} u$ fluctuate erratically, so we only show the running maximum. All of these errors are within or below $10^{-8}$. 

Next, we perform an independent residual test on our solution. That is, we evaluate the residuals of all equations of motion using finite differences and demonstrate convergence with increasing resolution.  By using spectral interpolation, we replace our spectral element mesh with uniform meshes at various resolutions, and we compute spatial derivatives via fourth-order finite differences.  Time derivatives are also computed using fourth-order finite differences, but we fix the time step.

\begin{figure}
\centering
\includegraphics[width=.45\textwidth]{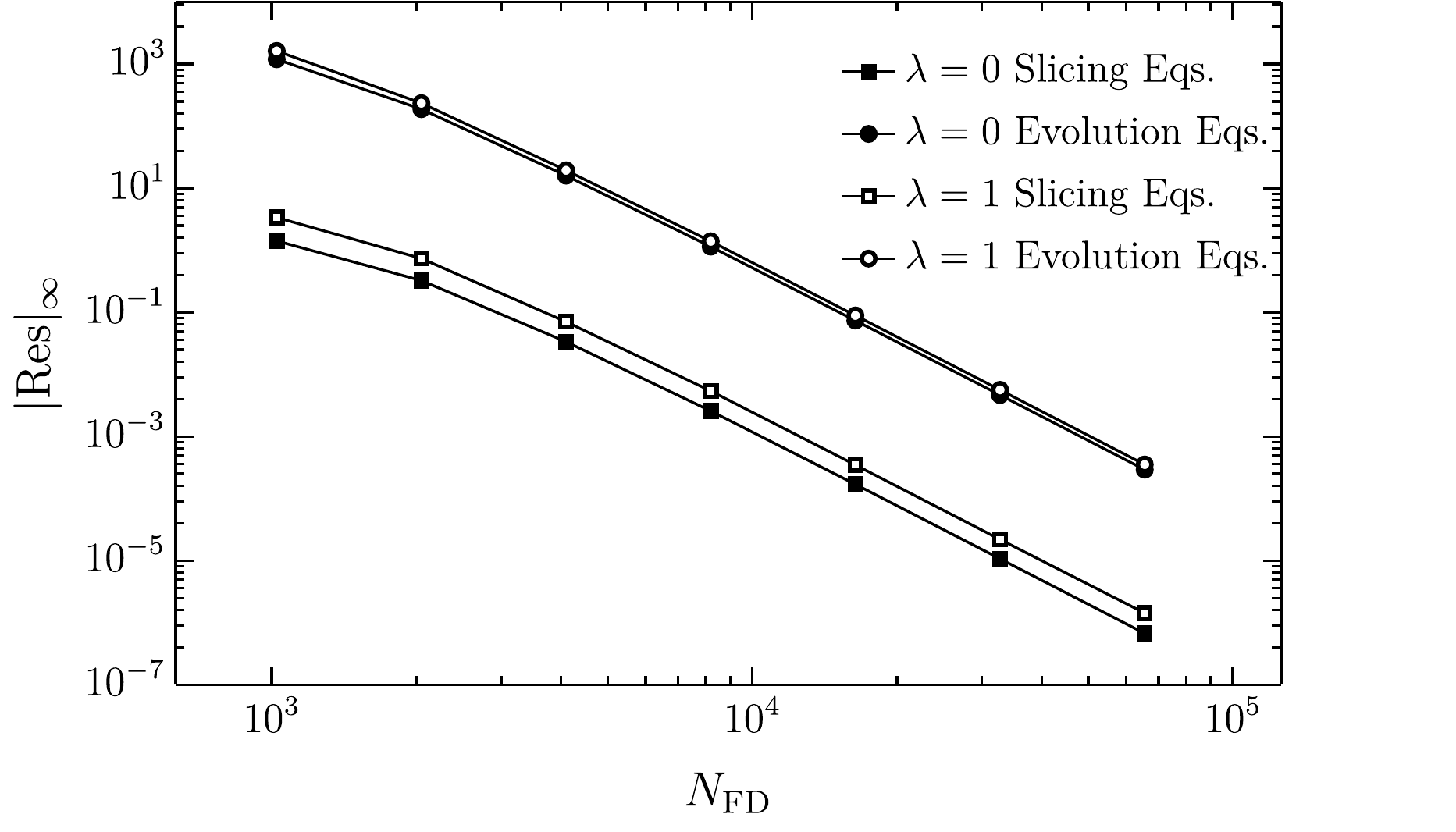}
\includegraphics[width=.45\textwidth]{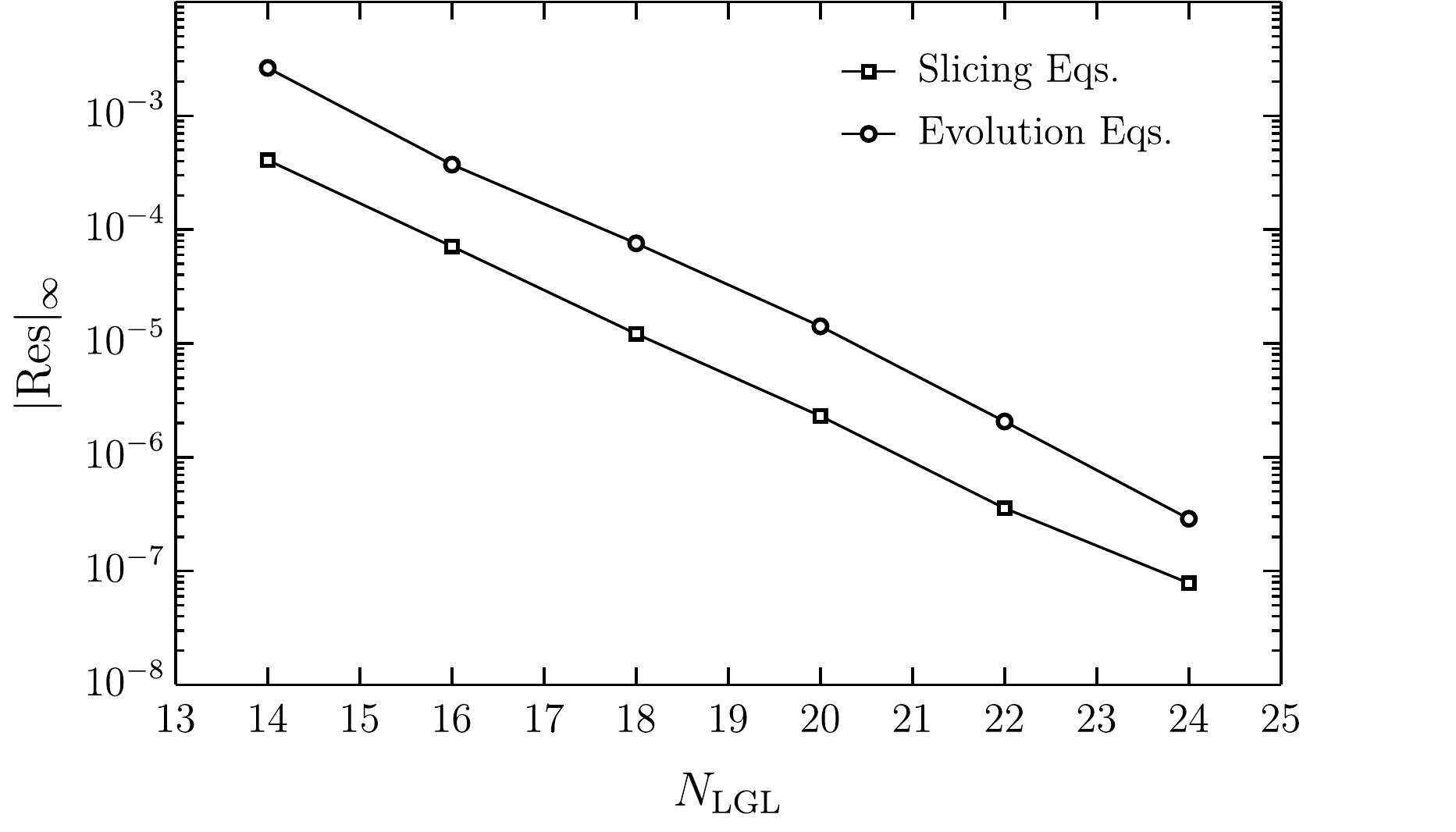}
\caption{\textit{Left:} Independent residual test using fourth-order finite differences on uniform grids with values interpolated from our spectral element mesh.  We have chosen a late time slice of a $\lambda=0$ run and a $\lambda=1$ run, and divided the 21 equations of motion into slicing equations and evolution equations. The power-law is consistent with fourth-order convergence.
\textit{Right:} Convergence of residuals at horizon formation for a perturbed boson star with $\omega=4.3$. We compute the evolution to horizon formation on a series of fixed meshes with 16 elements. We see exponential convergence by increasing the number of nodes within each element.
}\label{appendixfig:residualsconvergence}
\end{figure}  

We take a time step near collapse and compute the residuals of all 21 equations as described. The results are shown as a log-log plot in the left panel of Fig.~\ref{appendixfig:residualsconvergence}. We have divided the equations into slicing equations and evolution equations, and have taken the infinity norm of these sets of equations.  We find that the residuals converge with a fourth-order power law, as expected.

Finally, we test convergence of our code on a perturbed unstable boson star that leads to gravitational collapse.   We carry out this computation from start to horizon formation on a series of fixed meshes, each with 16 elements and varying numbers of (Legendre-Gauss-Lobatto) nodes within each element.  At the formation of an apparent horizon, we compute residuals as before. The right panel of Fig.~\ref{appendixfig:residualsconvergence} shows that the convergence is exponential, as expected of spectral methods.

\begin{center}  
{\bf Matching of Final States}  
\end{center} 

In Fig.~\ref{fig:finalstate}
of the main paper, we presented the evolution of the scalar response $|\langle\Pi\rangle|^2$ from two sets of collapsing initial data.  The scalar response $|\langle\Pi\rangle|^2$ in one of these vanishes at late times, which suggests that the final state is a Myers-Perry black hole.  In the other, $|\langle\Pi\rangle|^2$ approaches a constant at late times, suggesting that the final state is a hairy black hole \cite{Dias:2011at}.  In this section, we present further evidence of the final states by matching other quantities.  

One of these quantities is a pressure $\langle B\rangle$ extracted from the boundary stress tensor \cite{Balasubramanian:1999re,deHaro:2000xn} defined as
\begin{equation}
\langle B\rangle=2\pi G_5\left(\frac{1}{3}\langle T^{t}{}_{t}\rangle+\langle T^{\psi}{}_{\psi}\rangle\right)=\varphi_b(t,\rho=1)\;,
\end{equation}
where $\varphi_b$ is defined in \eqref{appendix:varphibdef}, and we have chosen a conformal frame where the boundary metric is $\mathbb R^{(t)}\times S^3$. For solutions within the symmetry class of our ansatz, the boundary stress tensor is fully specified by $\langle B\rangle$, $E$, and $J$. 

Once a horizon forms, we can also compute the entropy $S_\mathrm H$ and angular frequency $\Omega_\mathrm H$ of the apparent horizon.  Though these are not gauge-invariant quantities during time evolution, they must approach that of a stationary solution at late times. 

\begin{figure}
\centering
\includegraphics[width=.45\textwidth]{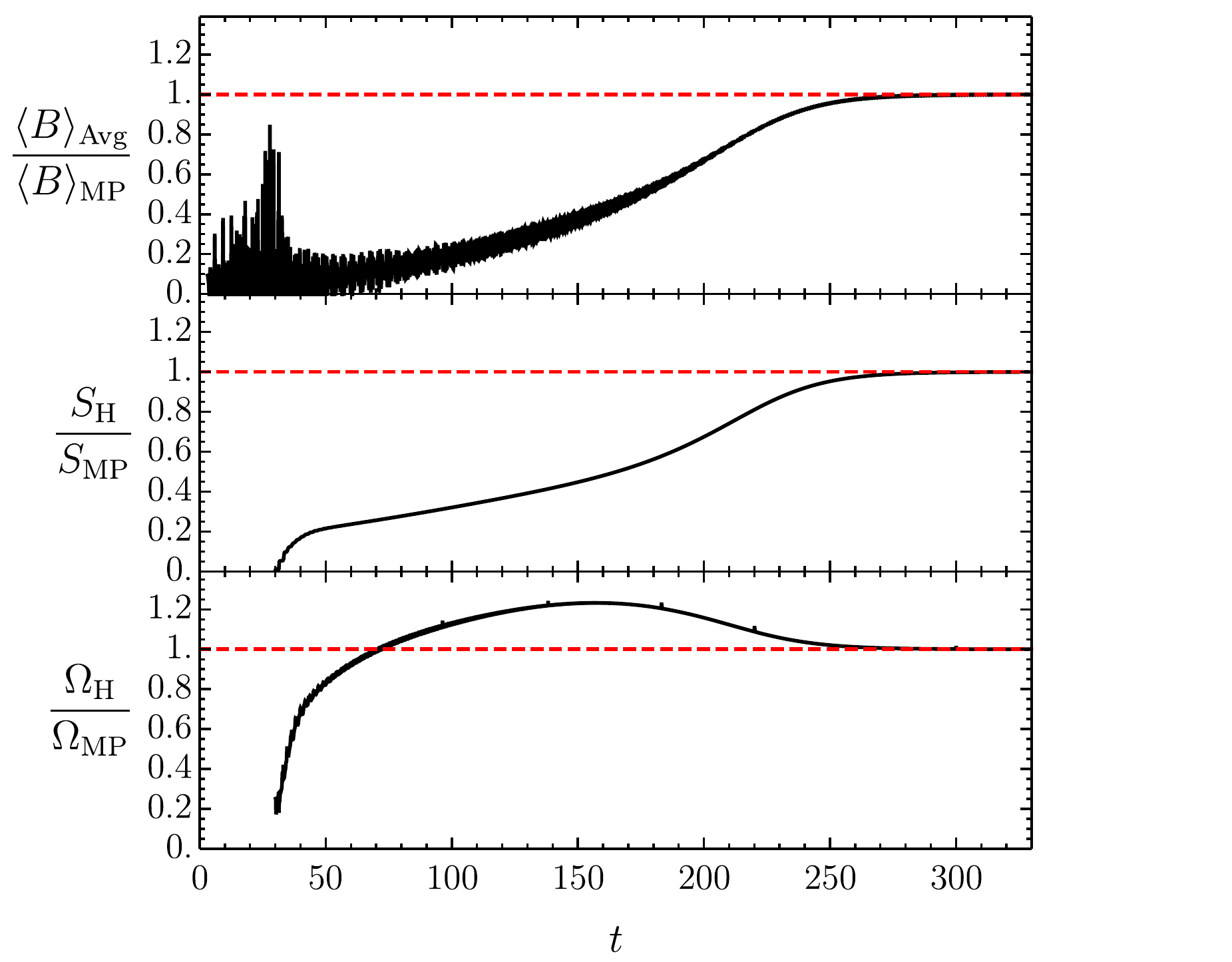}
\includegraphics[width=.45\textwidth]{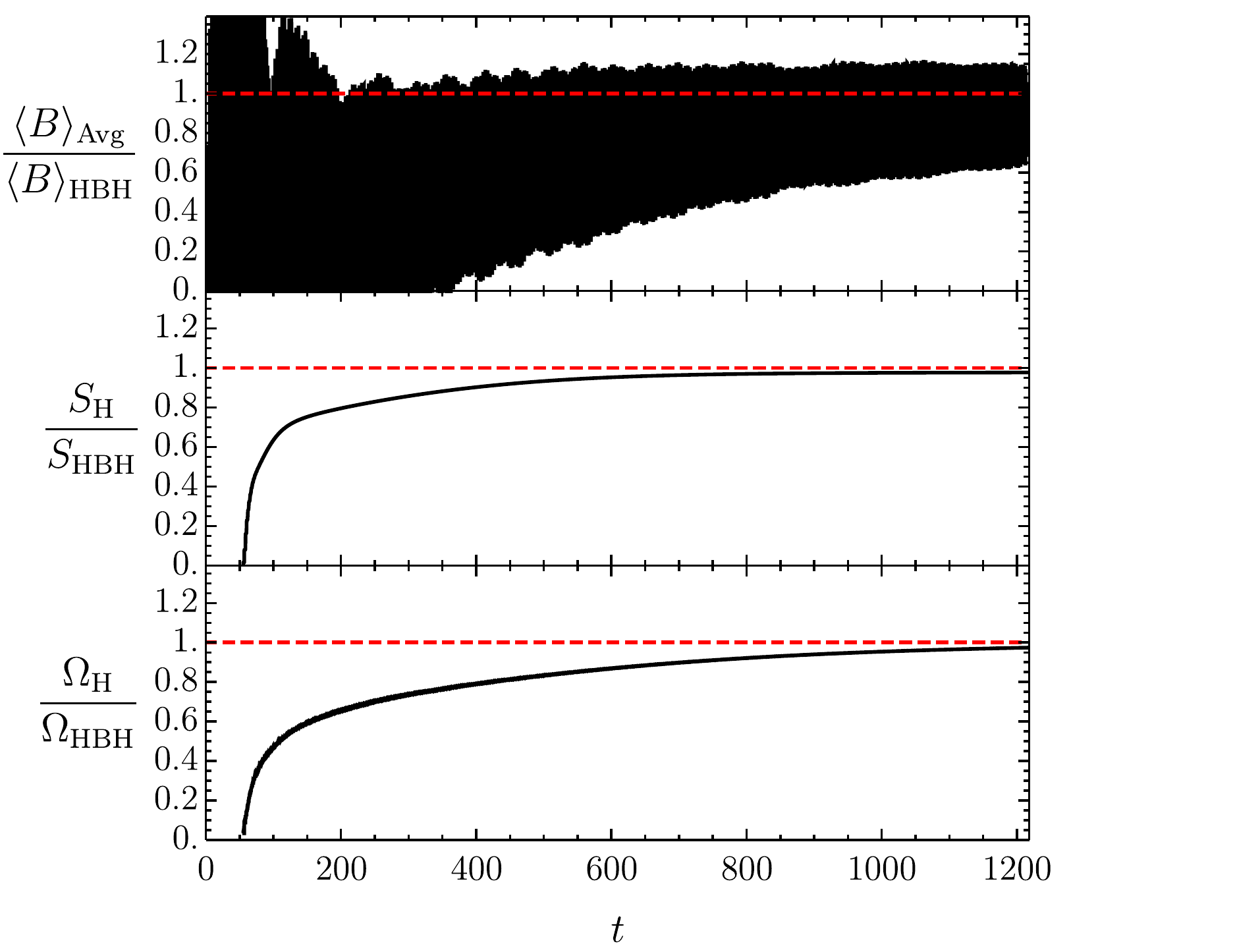}
\caption{Evolution of pressure $\langle B\rangle$ (averaged in a $2\pi$ time window), horizon entropy $S_\mathrm H$, and horizon angular frequency $\Omega_\mathrm H$ in units of their respective final stationary solutions. \textit{Left:} Evolution for subextremal Gaussian initial data with $\lambda=1$, $\epsilon=0.5$ (with $J\approx0.0136$, $E\approx0.0873$) that settles into a Myers-Perry black hole. \textit{Right:} Evolution for superextremal Gaussian initial data with $\lambda=1$, $\epsilon=0.4$ (with $J\approx0.00873$, $E\approx0.0560$) that settles into a hairy black hole.  The large oscillations in the top panel for $\langle B\rangle$ are a consequence of the small value for the corresponding hairy black hole solution $\langle B\rangle_\mathrm{HBH}\approx3.314\times10^{-5}.$}\label{appendixfig:matching}
\end{figure}

The evolution of these quantities is shown in Fig.~\ref{appendixfig:matching}, expressed in units of a stationary black hole solution \cite{Dias:2011at} with the same $E$ and $J$ as the initial data. We see that, at late times, these quantities approach that of the stationary black hole solution.  The final value for $\langle B\rangle$ in the superextremal case is the least conclusive of these since the remaining oscillations are still large compared to the hairy black hole value $\langle B\rangle_\mathrm{HBH}\approx3.314\times10^{-5}$.  The smallness of this value in this case makes matching difficult. 

\begin{table}
\centering
 \begin{tabular}{c|| c| c} 
  \text{QNM} & \text{Prony Analysis}  & \text{Perturbation Theory} \\  
 \hline
 $\langle B \rangle$ &$5.829-5.263\times 10^{-3}\,i$& $5.828-5.303\times 10^{-3}\,i$\\ 
 \hline
 $\langle \Pi_\mathfrak R \rangle$ & $4.7528-0.03130\,i$ & $4.7523-0.03127\,i$\\
 \hline
 $\langle \Pi_\mathfrak I \rangle$ & $4.7522-0.03126\,i$ & $4.7523-0.03127\,i$
\end{tabular}
\caption{Comparison of quasinormal modes (QNM) extracted using Prony analysis from late-time evolution of Gaussian initial data with $\lambda=1$, $\epsilon=0.5$ ($J\approx0.0136$, $E\approx0.0873$)  and from linear perturbation theory of the Myers-Perry black hole with the same $E$ and $J$.}
\label{appendix:table}
\end{table}

For the simulation that shares conserved quantities with a Myers-Perry black hole, we have also compared quasinormal modes. The late-time behaviour of these simulations should be well-approximated by linear perturbation theory of the final stationary state. Using Prony analysis, we can extract quasinormal modes from the time evolution of the scalar response $\langle \Pi_\mathfrak R\rangle$, $\langle \Pi_\mathfrak I\rangle$, as well as the pressure $\langle B\rangle$.  These can be compared to quasinormal modes computed directly from linear analysis of a stationary solution, which we have performed for Myers-Perry black holes.  The comparison is made in Table~\ref{appendix:table}, from which we see that the agreement is quite good.

\begin{center}  
{\bf Linear Perturbations of Boson Stars} 
\end{center} 

\begin{figure}
\centering
\includegraphics[width=.45\textwidth]{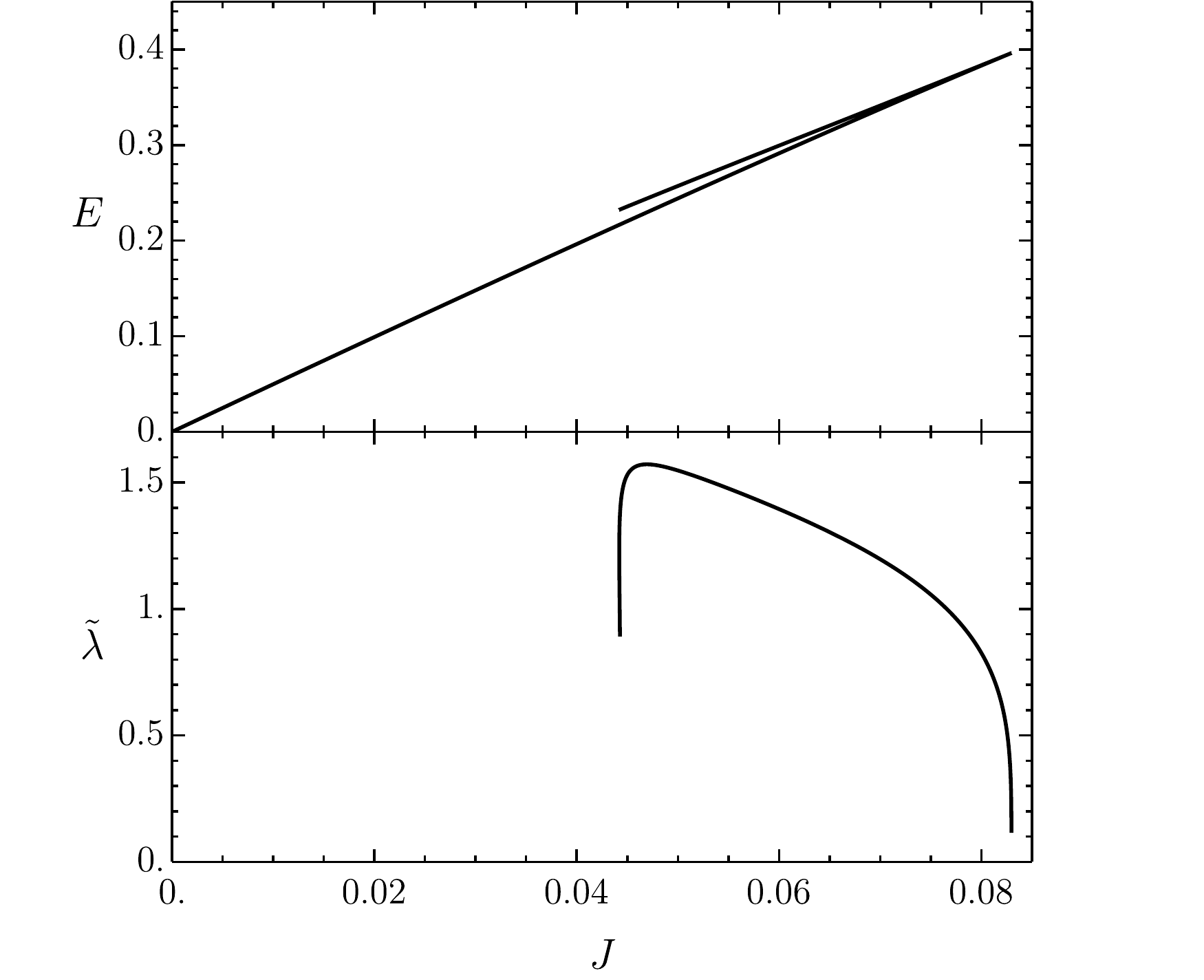}
\caption{\textit{Top:} Phase diagram of boson stars. Vacuum AdS lies at the origin with $E=J=0$.  \textit{Bottom:} Unstable mode for upper branch of boson stars.}\label{appendixfig:bstarmodes}
\end{figure}  

In this section, we demonstrate the linear instability of boson stars that lie beyond the turning point (see \cite{Dias:2011at} for details on this turning point, and the top panel of Fig.~\ref{appendixfig:bstarmodes} for a phase diagram).  One can typically prove that the existence of a turning point in the phase diagram implies an instability on at least one side of that point.  Studies of boson stars and related solutions in situations different from ours suggests that the side connected to AdS should be stable, and the other side unstable. Nevertheless, for completeness, we perform a linear stability analysis to confirm this expectation. 

For this purpose, we have decided to work in a different gauge than the rest of this paper.  We choose a Schwarzschild-like gauge with $\beta=0$ in the ansatz \label{appendix:metricansatz}. Boson stars are found in this ansatz by setting the metric to be time independent, and the scalar field to have a harmonic time dependence $\Pi(t,\rho)=e^{i\omega t}\tilde\Pi(\rho)$. The coupled gravity-matter equations for the boson stars are solved by Newton-Raphson iteration using pseudospectral methods on a Chebyshev grid (see \cite{Dias:2015nua} for a review on these methods). A seed is provided by perturbation theory about AdS. 

The functions are perturbed by choosing the form
\begin{equation}\label{appendix:pertform}
f(t,\rho)=f_0(\rho)+\delta f_c(\rho)\cosh(\tilde\lambda t)+\delta f_s(\rho)\sinh(\tilde\lambda t)\;,
\end{equation}
where $f$ stands for any metric or scalar field function, $f_0$ is the function on the boson star background solution, and the $\delta f$'s are their perturbations.  The equations of motion are expanded to linear order, yielding an eigenvalue problem with eigenvalue $\tilde\lambda$.  For the form we have chosen, real $\tilde\lambda$ is indicative of an instability.  There is a symmetry $f_s\leftrightarrow -f_s$, $\tilde\lambda\leftrightarrow -\tilde\lambda$, so without loss of generality, we take $\tilde\lambda$ to be positive. Each eigenvalue will have a multiplicity of two.  This degeneracy arises because one can shift time by a phase $t\to t+\phi_0$ and preserve the form of the functions \eqref{appendix:pertform} above.

We again employ pseudospectral methods to solve the eigenvalue problem \cite{Dias:2015nua}.  We first solve the resulting matrix (generalized) eigenvalue problem by QZ factorization.  We find no unstable modes in the branch of boson stars connected to AdS.  Past the turning point, we identify an unstable mode. We track this mode in parameter space using Newton-Raphson iteration. To eliminate the degeneracy, we demand (in addition to a normalisation condition) that one of the perturbation functions takes a certain value at a spatial point.  Since the resulting matrix problem is overconstrained, we solve it by linear least squares. We have checked that the solutions remain the same with different numerical resolutions.

In Fig.~\ref{appendixfig:bstarmodes}, we show the phase diagram of boson stars and the eigenvalue corresponding to the unstable mode. We see that the eigenvalue approaches zero near the turning point (around $J\approx 0.083$), suggesting that the turning point is indeed the onset of this instability.  

Interestingly, this eigenvalue also decreases rapidly near the edge of our numerics around $J\approx 0.045$.  Near this region of parameter space, the phase diagram has a number of further turning points (and plots of $E$ and $J$ versus the boson star frequency $\omega$ produce spirals). This opens the possibility that boson stars may stabilize again past another turning point. Similar behaviour has been conjectured in \cite{Figueras:2012xj} in the context of non-uniform black strings. However, it is difficult to proceed further since the solution approaches a singularity. 

\bibliography{refs}{}
\end{document}